# MOTIVATION FOR HYPERLINK CREATION USING INTER-PAGE RELATIONSHIPS

Patrick Kenekayoro[1]   Kevan Buckley[2]   Mike Thelwall[3]

[1] *Patrick.Kenekayoro@wlv.ac.uk*   [2] *K.A.Buckley@wlv.ac.uk*   [3] *M.Thelwall@wlv.ac.uk*
Statistical Cybermetrics Research Group, University of Wolverhampton, Wulfruna St
Wolverhampton, West Midlands WV1 1LY  (UK)

**Abstract**
Using raw hyperlink counts for webometrics research has been shown to be unreliable and researchers have looked for alternatives. One alternative is classifying hyperlinks in a website based on the motivation behind the hyperlink creation. The method used for this type of classification involves manually visiting a webpage and then classifying individual links on the webpage. This is time consuming, making it infeasible for large scale studies. This paper speeds up the classification of hyperlinks in UK academic websites by using a machine learning technique, decision tree induction, to group web pages found in UK academic websites into one of eight categories and then infer the motivation for the creation of a hyperlink in a webpage based on the linking pattern of the category the webpage belongs to.

## Introduction

Webometrics has been defined as "*the study of web based content with primarily quantitative research methods for social science goals using techniques that are not specific to one field of study*" (Thelwall, 2009).  Techniques from different fields like mathematics and statistics have been applied to study web content for webometrics research. Machine learning is an area in computer science that is concerned with pattern discovery. This technique has not been used extensively in webometrics research, but has been applied in several computing web studies, for example (Chau & Chen, 2008; Luo, Lin, Xiong, Zhao & Shi, 2009; Qi & Davison, 2009).

A particular area where machine learning can be applied to webometrics research is in link analysis. Link analysis involves the study of link relationships between a group of websites or the link structure of a group of websites. It has been successfully used as a source of business intelligence (Vaughan & Wu, 2004; Vaughan, 2005) and used in several studies of academic websites (Thelwall, 2002c; Thelwall & Wilkinson, 2003).  Nevertheless, using raw link counts between websites can be unreliable and several researchers have attempted to find alternatives to raw link counting or tried to understand the meaning of link counts between websites. Researchers have classified links in the web pages of academic institutions as research or non-research related (Thelwall, 2001), substantive or non-substantive (Smith, 2003) and shallow or deep (Vaseleiadou & van den Besselaar, 2006). Other researchers have tried to identify the reasons why links in academic web pages were created (Bar-Ilan, 2004; Wilkinson, Harries, Thelwall, & Price, 2003) but there is no agreement about an effective way to classify the reasons for university interlinking on a large scale, which is the goal of this paper. The manual classification of individual links is time consuming, thus it is infeasible for large scale studies. Perhaps this is the reason why there is a dearth of literature in link classification. In this paper, the reason for link creation is inferred from the relationship between the two pages the hyperlink connects; the source page and the target page. Hyperlinks have been previously classified using the target page (Thelwall, 2001), using both the source and target page can give better insight to the reason for hyperlink creation and web page classification is a simpler problem than individual link classification.

Typically, a university's website has thousands of web pages so an effective approach will be to group similar web pages together and then infer why a link is created based on the link creation motivation of the group that web page belongs to. This makes it necessary to identify the types of web pages that can be found in a university's website. Page types are identified using the mission of a university and the function of its website as a guide.

Stuart, Thelwall and Harries (2007) suggest that methods for automatic classification of hyperlinks should be developed if links are to be fully harnessed for webometrics research. The reason why a link is created, that is the relationship between the source and target page can form classes of links in a university's website. Machine learning techniques are used to automate the classification scheme, there by bringing us a step closer to fully harnessing hyperlinks for webometrics research.

The main goal of this study is to identify a method that effectively determines the reason why a link in a UK university's website has been created. This is achieved by grouping web pages into categories that are in line with the three missions of Higher Education Institutions (HEIs), automating the classification scheme with machine learning techniques and then examining the relationship between page categories. This paper begins with background information on link classification, the supervised learning technique used in this study, then, categories of pages that could be found in UK university's websites are identified, and the results of classification with a supervised learning technique are shown and the relationship between a random sample of web pages analysed.

## Link Classification

Webometrics can be used as a source of business intelligence. Vaughan and Wu (2004) showed that a link count to a company's website positively correlates with the company's business performance and co-linked web pages were used to identify a company's competitors (Vaughan, 2005). Webometrics has also been applied in the study of academic institutions (Payne & Thelwall, 2004; Thelwall

& Wilkinson, 2004; Vaughan & Thelwall, 2005) and in the identification of political trends (Park & Thelwall, 2008; Romero-Frías & Vaughan, 2010; Romero-Frías & Vaughan, 2012).

Raw link counts are unreliable (Thelwall, 2002a; Thelwall, 2002b) because links are prone to spamming (Smith, 1999) and there could be different motivations behind the creation of hyperlinks (Wilkinson et al., 2003). There are different reasons behind link creation, it can vary according to the function and operational relationship between the different organisations studied (Minguillo & Thelwall, 2011). Linking between municipalities in Finland is motivated by cooperation made possible because of geographic closeness (Holmberg & Thelwall, 2009) and co-linking of business web pages tends to be for business reasons (Vaughan, Kipp, & Gao, 2007) which is different from co linking of university web pages that could just be as a result of general reference. This difference highlights the disparity between linking patterns in different domains, thus caution should be used when applying a method tested in one domain to a different domain.

Several attempts have been made to classify web pages in a University's website but there is no consensus as to how links can be classified and Thelwall (2006) suggests no single link interpretation is perfect. Two approaches to hyperlink interpretation are (Thelwall, 2006)

- Interviewing a random selection of link creators about why they created a link
- Classification of a random selection of links in a way that is helpful to the research goals

Author interviewing may give a more accurate result but classification of links is a more practical approach; it is the most common method for hyperlink classification in the academic web space (Bar-Ilan, 2004; Bar-Ilan, 2005; Thelwall, 2001; Thelwall, 2003; Wilkinson et al., 2003).

Thelwall (2001) classified hyperlinks in web pages of UK academic institutions as research related or not research related based on the content of the target page, which he noted was a practical step. Although classification of some pages was subjective, a situation similar for all research in this area, some general rules were created to for the classification process. For example, departments' homepages, staff research profiles, web pages of research groups were classified as research related while electronic journal pages were classified as non-research related. Results showed that using only research related links increased the correlation with average research rating of UK institutions.

Wilkinson and his colleagues (2003) studied 414 random links between UK academic institutions in order to identify the motivations for hyperlink creation. Even though individual links were investigated, the reason from link creation was determined using the source page and target page. They suggest that this approach is difficult as it is impossible to guess the motivation for link creation and in some cases there could be several motivations.

Thelwall (2003) studied 100 random inter site links from a UK university's' web page to the homepage of another UK university. He grouped web pages into four categories: *navigational*: a link created to direct uses to other non-subject specific information, *ownership*: links to partners and they were often in the form of a clickable image of the university's crest, *social*: links to institutions of collaborating research groups and *gratuitous*: links created without any specific motivation.

Bar-Ilan (2004), in perhaps the most systematic study so far, classified the link, source page and target page from different aspects, link context, link tone and several other properties, in a case study of eight universities in Israel. This approach is difficult as Wilkinson and his colleagues (2003) pointed out problems with guessing author motivations and subjective decisions in the classification process. It is also impractical to study each link individually because of the sheer number of links that could be found in a university's website. A more practical approach to link classification is finding the relationship between the two pages a hyperlink connects, the source and target page, which reduces the problem of hyperlink classification to web page classification. To study the relationship between web pages, the type of pages that could be found in a university's website must be identified.

**Supervised Learning**

Although web page classification is simpler than individual link classification, this process is still infeasible for large scale manual studies because a typical university's website can have thousands of web pages.

Supervised machine learning is a Computer Science technique concerned with teaching machines to predict unseen cases of input data based on patterns identified from previously observed examples, usually called a training set. There are several machine learning algorithms, like decision tree induction, support vector machines, neural networks and k nearest neighbours' classifiers. Decision tree induction has an advantage over other models in that it is easy for humans to understand the resulting classifier, because of its high level rules, as opposed to other black box models like neural networks whose model is encapsulated in a complex numerical model. For this reason, decision tree induction was used in this study, but other classification techniques may produces similar or better results.

*Decision tree induction*

Decision tree induction recursively splits data into disjoint sets according to a criterion. Each node is a feature an instance can have, and leaf nodes contain output classes for instances that reach that node. Figure 1 is an example of a decision tree classifier that classifies instances into one of three possible classes. Classification of instances start at the root node, then the instances traverse down the tree in the direction that meets several criteria until a leaf node is reached. The value of the leaf node is then assigned to that instance.

Constructing optimal binary decision trees is a NP-complete problem (Kotsiantis, Zaharakis, & Pintelas, 2006), however several techniques like the C4.5 algorithm (Quinlan, 1993) and CART, acronym for Classification And Regression Trees (Breiman, Friedman, Stone, & Olshen, 1984) can be used to build decision trees. CART and C4.5 algorithms are implemented in a machine learning toolkit, WEKA (Hall et al., 2009) that is used in this study to automate the classification scheme.

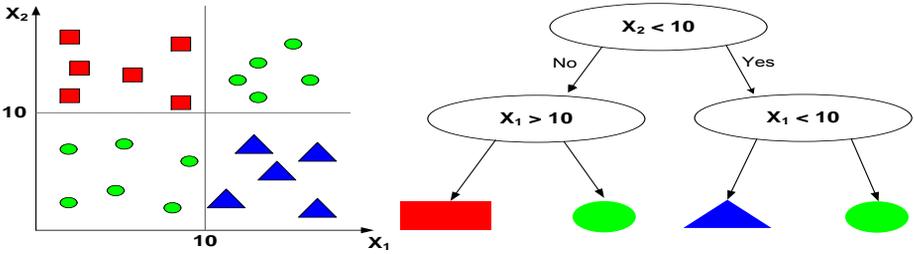

**Figure 1 A Decision Tree Classifier**

Two major phases of a decision tree induction are the growth phase and the pruning phase (Kotsiantis, 2011). The growth phase involves splitting the training data into disjoint sets and the pruning phase reduces the size of the decision tree to avoid overfitting.

**Decision tree induction Pseudo Code** (Kotsiantis, 2007)
1. `Check for base cases`
2. `For each attribute a`
3. `   Find the feature that best divides the training data`
4. `Let a_best be the attribute that best splits data`
5. `Create a decision node that splits on a_best`
6. `Recurse on the sub-lists obtained by splitting on a_best and add nodes as children`

A major difference between the C4.5 and CART algorithms is the way the best feature that separates the training data is selected. The attribute with maximum information gain is used to split the training set. C4.5 uses entropy to compute the information gain, while CART uses the Gini index. The entropy is calculated by:

$$Entropy(S) = - \sum_{i=1}^{n} Freq(C_i, S) * log(Freq(C_i, S))$$

$Freq(C_i, S)$ is the relative frequency of instances in class $C_i$.

The Gini index is computed by:

$$GiniIndex(S) = 1 - \sum_{i=1}^{n} Freq(C_i, S)^2$$

And information gain is computed by:

$$InformationGain(S, A) = I(S) - \sum_{i=1}^{n} \frac{|S_i|}{|S|} * I(S_i)$$

The formula above computes the information gain of attribute A in data set S where $i = 1 \ldots n$ are possible values of attribute A, $S_i \ldots S_n$ are partitioned subsets of S where attribute A is $i$, $I(S)$ is the entropy in C4.5 algorithm and Gini index in CART algorithm.

*Testing and Evaluation*

It is essential to evaluate the accuracy of a learning algorithm. The fundamental assumption of machine learning is that the distribution of training data is identical to the distribution of test data and future examples (Liu, 2006). If the learning algorithm generalizes the training set, then the machine learning assumption suggests that it will perform well for future unseen examples. Generalization is estimated by the accuracy of the learning algorithm, measured by the equation:

$$Accuracy = \frac{Number\ of\ correctly\ classifed\ examples}{Total\ number\ of\ examples}$$

Ultimately, the accuracy measure depends on the application which applies the learning model. Precision, recall and F-measure give a more elaborate description of the performance of a learning algorithm. They are calculated based on four parameters: True Positives (TP), False Negative (FN), False Positive (FP) and True Negative (TN). Given the test set D, if each instance of the set can be of class $y = (1, -1)$ and $f(x)$ is the function trained to predict future unseen instances. The parameters TP, FN, FP and TN are:
- True Positives   (TP)  $= Count\ (f(x) = 1\ and\ y = 1)$
- False Negatives  (FN)  $= Count\ (f(x) = -1\ and\ y = 1)$
- False Positives  (FP)  $= Count\ (f(x) = 1\ and\ y = -1)$
- True Negatives   (TN)  $= Count\ (f(x) = -1\ and\ y = -1)$

Precision, recall and F-measure is computed by:

$$Precision = \frac{TP}{TP + FP}\ ;\quad Recall = \frac{TP}{TP + FN}\ ;\quad F_{measure} = \frac{2 * Precision * Recall}{Precision + Recall}$$

Precision and recall are used when the interest is on one particular class. In most cases, the accuracy formula earlier defined as percentage of correctly classified divided by the total number of instances is used.

A rule of thumb in machine learning is that the input data should be divided into two-thirds for training and the remaining one-third for tests/validation. Another technique, cross validation is also used. In cross validation, the input data is divided into $n$ equal disjoint subsets. Each subset is used as the test set, and the union of the others the training set. The accuracy of the classifier is the average accuracy of the $n$ different subsets. A special case of cross validation is the leave one out approach, where a single example is used as test and all others for training. In this case, $n$ is the number of training examples, thus this method can be computationally expensive.

**Page Types**

Bar-Ilan (2005) created a detailed framework for characterising the uses of links in an academic website. The scheme had 31 possible relationships between two pages and 13 different page types. Using machine learning to automate this classification scheme can be difficult because of the number of variables involved. Moreover, this study aims to group web pages based on the three missions of Higher Education Institutions (HEI) and the functions of its website. Bar-Ilan's (2005) method does not fit into this classification scheme. For example, a physical unit can comprise of an administrative unit or a research unit. These two units serve different purposes in regard to the missions of HEIs so they should not be grouped together. Because of this, this study uses a classification scheme that is less detailed than (Bar-Ilan, 2005), but is easier to automate with machine learning, and is more in line with the aims of this paper.

If the links in a university's websites are to be classified based on the relationship between two pages, it is necessary to identify the types of pages that can be found in a university's website, and then study how these page types interlink.

In order to identify the type of pages in a university's website a random set of web pages in a university's domain were visited and manually classified. A custom web crawler was designed to get the link structure of 111 UK universities. The crawler extracted links originating from a UK university to another UK university, not visiting all pages in a university's website. It only covers links that can be reached by following links from a university's homepage, similar to Thelwall (2001). An additional constraint was added as this work is only concerned with hyperlinks between UK academic institutions. New web pages were not added to the list of websites to visit when the crawler visited 2000 consecutive pages without finding a link to another UK university. 15 link pairs between universities were randomly selected and then the web pages that these links direct to were used to identify the page types in a university's website.

Websites of organisations are designed to disseminate the activities and functions of that organisation. In some cases the structure of the website replicates the physical structure of that organisation. Nowadays, Higher Education Institutions (HEIs) have three main goals, teaching, research and what authors simply refer to as the third mission. If websites of universities are designed to channel the activities and functions of that university, which in turn are in-line with the (three) main goals of HEIs then university websites will be similar. From a university's homepage a general idea about the goals of the university as well as the function of its website can be inferred. Thus text from the homepages of UK universities was used to determine the types of pages studied in this work. Text from the homepages of UK universities was extracted, and then the top 20 words when stop words were removed were used as a guide to determine preliminary page types.

**Table 1 Page Type and description**

| Page Type | Description |
|---|---|
| About | Promotes the school and gives information to staff/students. Examples of such pages are news, information, university profile, prospectus, events |
| Business and innovation | Connects the school to non-academic environment. Examples include expert services offered, community projects, partnership with science parks |
| Discussion | Forums, blogs or web page containing opinions of a user. Comments or posts in these pages are for a variety of reasons; research, teaching or recreational. |
| Support | Contains a repository of learning resources for students/staff support, skills for learning, services, counselling. Examples include Archives, Books, Database. |
| Research | Involved with the production of new knowledge. Examples include Research centres, research groups, research projects, academic schools/departments, conferences, Abstract, Academic Article |
| Staff | Related to a staff in the university. Examples include staff homepage, staff profiles, list of publication, CV |
| Student Life | Enhances the student experience. Examples include student union website, student benefits, campus facilities, tourism, recreation |
| Study | Involved with transfer of knowledge. Examples include module learning materials, module timetables, module page, lectures |

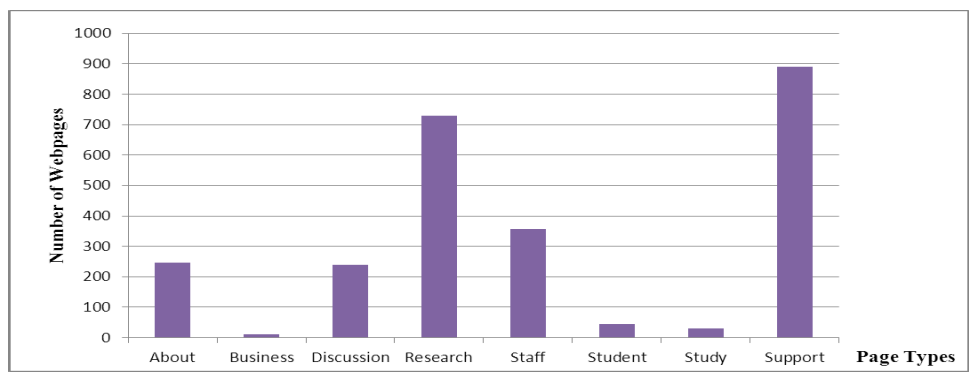

**Figure 2 Distribution of page types in 2500 random UK university web pages**

100 random web pages were given to an independent researcher for classification. When a page type identified by the independent researcher did not fit into the preliminary page types, a new category was created.

The page types in Table 1 are largely based on the authors' opinion; top words were only used to assist in the decision process. These page types however, cover the majority of web pages found in university's websites. Pages were grouped into one of the eight categories and then the possible reasons for link creation were identified for links originating from one page type to another.

**Automatic Classification**

Decision trees are constructed using features of the training set. Different features may associate a training instance to a particular page type. In this case, training instances are the web pages to be classified. 2500 web pages were randomly selected and manually classified into one of eight categories a UK university's web page could belong to. Two thirds of the web pages were used for training and the rest for testing.

The features of each web page were derived from the web page title and/or web page URL pre-processed and then represented as a word vector of TF (term frequencies) or inverse document frequency multiplied by the term frequency (TFIDF). In this case, a term frequency representation is similar to a binary representation because page titles are short thus words rarely occur more than once.

Pre-processing transforms the text to an information rich representation. Pre-processing steps used are:

**Word Tokenization**: Splits attributes (web page title/URL) into word tokens. For example "the quick brown fox jumps over the lazy dog" has 9 (nine) tokens of 8 (eight) word types. A simple way to achieve tokenization is by assuming [space] separates word tokens. Only words that did not contain any non-alphabetic characters were used in this study.

**Capitalization**: All characters were represented in lower case.

**Stop word removal**: Removal of the most frequent words that occur in the English language. Words like "the, and, a, is ... " are all removed. WEKA (Hall et al., 2009) contains the list of stop words that was used in this study. The 111 university names as well as www, http and https were added to the list of stop words.

**Stemming**: Stemming reduces inflected words to their root form or stem. For example jumps and jumping have the same stem, jump. Accuracy may be improved if all words are represented in their root form. The Porter Stemmer is a commonly used stemming algorithm and it is used in this study. Stemming algorithms occasionally make errors. For example, the Porter Stemmer stems both university and universe to univers.

WEKA (Hall et al., 2009) implements decision tree induction in its J48 algorithm. Although the default settings may give satisfactory results, in some cases tweaking the settings may improve the accuracy of the algorithm. Feature selection as well as data pre-processing is also an important aspect in supervised learning. Table 2 shows how pre-processing options influence the accuracy of the classifier. Training accuracy is determined using a 10 fold cross validation, while verification is determined using the formula $Verification = \frac{Number\ of\ correctly\ classified\ examples}{Total\ number\ of\ examples}$; the data set used in verification is different from the set used in training. Verification gives an estimate of the out of sample performance of the learning algorithm and is also used to identify overfitting. A machine learning algorithm overfits when it performs better in training than in testing.

**Table 2 Training and Verification of top 10 pre-processing options of decision tree induction**

| Bigrams/ Unigrams | TF *IDF | Stem | Stop words | Page title | URL | Training | Verification |
|---|---|---|---|---|---|---|---|
| Unigrams | | Yes | Yes | Yes | Yes | 72.13 | 71.25 |
| Unigrams | Yes | Yes | Yes | Yes | Yes | 72.16 | 71.25 |
| Unigrams | Yes | Yes | | Yes | Yes | 71.15 | 69.9 |
| Unigrams | | Yes | | Yes | Yes | 71.16 | 69.9 |
| Bigrams + Unigrams | | | Yes | Yes | Yes | 71.57 | 69.66 |
| Bigrams + Unigrams | Yes | Yes | Yes | Yes | Yes | 72.31 | 69.66 |
| Bigrams + Unigrams | | Yes | Yes | Yes | Yes | 72.32 | 69.65 |
| Unigrams | | | | Yes | Yes | 71.86 | 68.92 |
| Unigrams | Yes | | | Yes | Yes | 72.66 | 68.92 |
| Unigrams | | | Yes | Yes | Yes | 72.78 | 68.55 |

Settings of the classifier were tweaked and the best results are shown in Table 2. On average, the top 250 features were used in the construction of the decision tree.

When word counts are used as features, the number of features increases as the training set increases. Too many irrelevant features affect the speed as well as accuracy of a learning algorithm, so input features have to be carefully selected. The J48 algorithm uses the information gain to select the best feature that optimally splits the training data, so it is logical to use information gain to identify relevant features. Other methods like principal component analysis and entropy-weighted genetic algorithm can also be used to reduce the size of the features. Another way to reduce the feature size is by generation an initial decision tree using all features, and then excluding those features not used in the initial tree during subsequent training. In tests, only 7 percent of the features were used in the final decision tree. Excluding features that were not used produced a slight improvement in the accuracy of the decision tree.

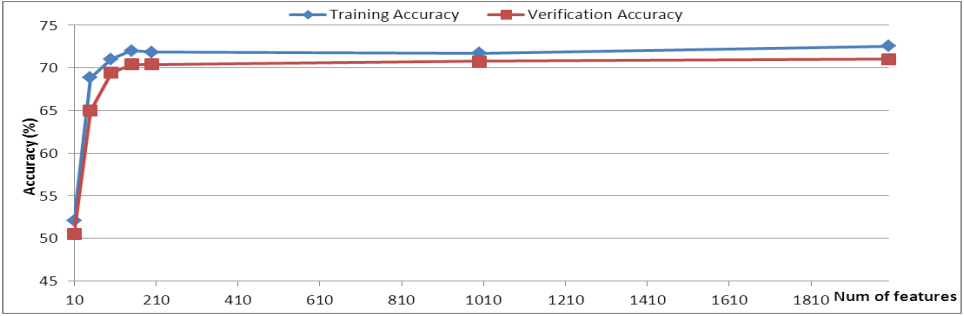

**Figure 3 Influence of feature size on accuracy of the classifier**

Figure 3 shows how the number of features affects the accuracy of the decision tree classifier when the best pre-processing setting from Table 2 was tested for different number of input features. Initially, as the feature size increases, the accuracy of the classifier also increases but at some point the increase in size doesn't improve the accuracy; it only reduces the speed of the decision tree classifier.

**Table 3 Classification Accuracy of each page type**

| Class | Precision | Recall | F Measure |
|---|---|---|---|
| About | 0.59 | 0.46 | 0.52 |
| Business and Innovation | 0.00 | 0.00 | 0.00 |
| Discussion | 0.87 | 0.89 | 0.88 |
| Research | 0.63 | 0.80 | 0.70 |
| Staff | 0.78 | 0.75 | 0.77 |
| Student Life | 0.63 | 0.29 | 0.40 |
| Study | 1.00 | 0.33 | 0.50 |
| Support | 0.78 | 0.71 | 0.74 |

The overall accuracy of the decision tree classifier is about 71%. However it is interesting to know how accurately the classifier identifies individual page types. This is determined using precision, recall and F measure in Table 3. In summary, precision is the likelihood that the classifier will correctly classify a web page of type X as class X, while recall is the likelihood that the classifier will not classify a web page that is not of type X as class X. F measure is the accuracy of an individual class computed by a formula that depends on the precision and recall.

Figure 4 shows a partial decision tree for the classification of the web pages. This is not the optimal result that can be achieved. The settings of the classifier were adjusted to reduce the size of the tree. If the decision tree in Figure 4 is used to classify university web pages, several page types will always be incorrectly classified. Business and Innovation, Study and Student Life pages do not appear in any leaf node so they will always be misclassified. Web pages that do not contain any of the keywords will be classified as research pages. The tree in Figure 4 has an accuracy of 46.8%.

The nodes in the tree above show top terms in the feature set that are associated with a specific page type.

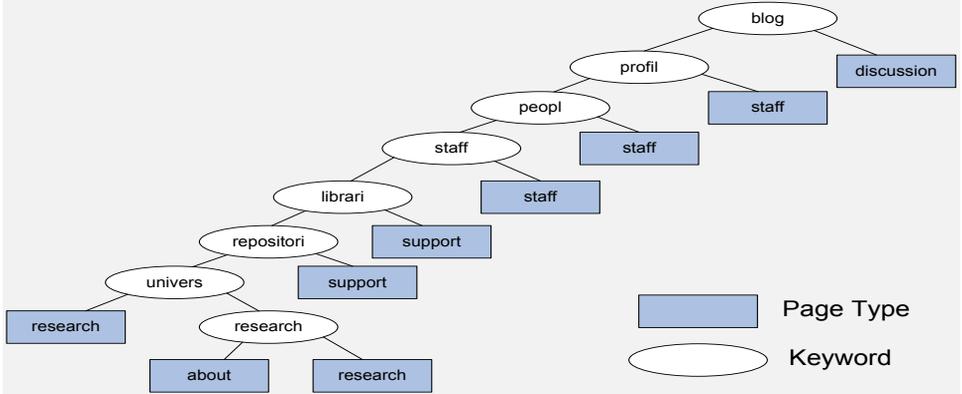

**Figure 4 Tree view of a decision tree induction classifier**

### Inter-page Relationships

Each page type was studied to identify the type of pages they link to and possible reasons for interlinking. 15 random links were randomly selected but at the time of the investigation, some of these pages, either the source or target pages were not available online. Such links were excluded from the study.

With eight page categories, if each category has a link to all other categories including itself there becomes $8 * 8 = 64$ page relationships to study. However, not all page types interlink and majority of links are between the same page types. Subsequent sections describe results for each page type.

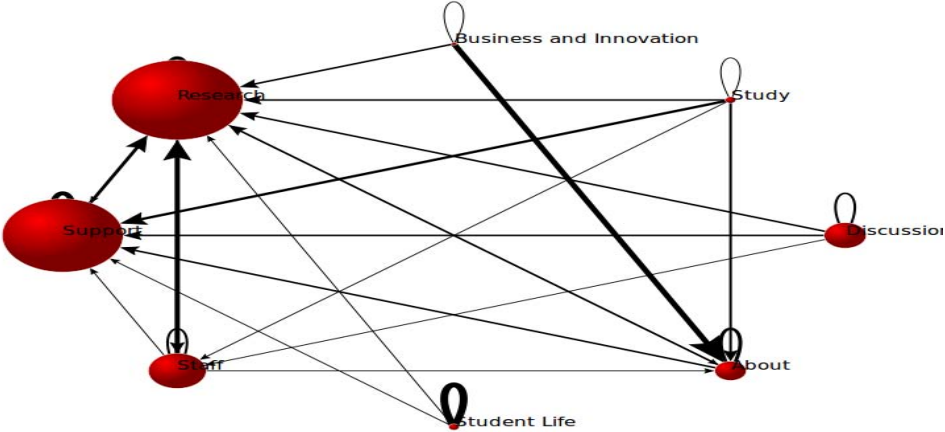

**Figure 5 Visual representation of interlinking between page types**

**Table 4 Page types and reasons for link creation**

| Page Type | Size | Additional Notes |
|---|---|---|
| Support | 35% | ❖ Rarely link to other page types<br>❖ Links to research pages that own or created the resource in the support page.<br>❖ Links are created to direct users to other relevant information, often to other pages that are created to improve learning, research or teaching skills. |
| Research | 28.6% | ❖ Links to About pages, usually a clickable logo of a collaborating university; organisational links as described by Thelwall (2003).<br>❖ Pages about research projects had links to staff pages of its collaborators or homepages of research groups or department.<br>❖ Research pages had links to all research groups or departments in the same scientific field. |
| Staff | 14% | ❖ Links to about pages (homepages of universities) were often what Thelwall (2003) refers to a gratuitous links.<br>❖ Links to support pages that contain a resource, for example, staffs' publications.<br>❖ Links to other staff pages because of collaboration in a research project or co-authorship in a publication. |
| About | 9.7% | ❖ Are linked to but rarely link to other universities.<br>❖ Largely made up of course prospectus and university homepages.<br>❖ Majority of outgoing links are for non-scholarly reasons. |
| Discussion | 9.4% | ❖ Links are created for a variety of reasons, so it is very difficult to identify a general pattern.<br>❖ Each blog entry belongs to a particular page type, and reasons for linking are the same as reasons of its corresponding page type. |
| Student Life (SL) | 1.8% | ❖ Mainly Link to SL pages in close geographic locations.<br>❖ A part of the reason why link analysis research shows that UK universities links to other universities in close geographic location. |
| Study | 1.2% | ❖ Majority of links are to support pages containing information relevant to the course.<br>❖ Links to research pages that contained software/ research output used in the course.<br>❖ Links to staff pages that authored course material, or a visiting professor<br>❖ Represents only a small set of total pages, perhaps because teaching materials are located on a protected server, thus inaccessible through public web crawlers. |

Figure 5 shows how different page types inter links. Vertices represent page types and arcs represent links from a page type to another page type. The size of the vertices indicates the number of web pages in that page type, while the colour of arcs indicates the percentage of link from a page type to another page type.

Thicker arcs mean a large percentage of links from that page types go to the page type on the other end of the arc, while bigger vertices mean a large percentage of web pages belong to this page type.

Random links are manually classified in order to identify linking patterns of different page types. Table 4 shows linking patterns identified for different page types.

**Conclusion and Further Work**

Hyperlink classification based on inter-page relationships is a practical solution compared to other methods that try to classify individual links on web pages. This work used the source and target page to determine the reason for linking. The relationship between the source and target pages was similar for web pages in the same category.

Support pages had links to its resource creator or pages that gave additional information that enhances teaching or learning schools. Staff web pages about a research project are ideal to identify collaboration or cooperation between institutions. Other research pages only show the amount of research activity going on in the university. Web pages about the living experience in the university, even if they represent only a small part link to web pages in close geographical regions. These links are for non-scholarly reasons and should be excluded when identifying academic relationships between universities. Other page types, business and innovation and study web pages contributes less than 1 percent to the total links to other universities, perhaps because they are situated in an area not accessible by web crawlers or link to other non-academic originations. Administrative pages also contained few links to other universities, and they were for non-scholarly reasons. However, majority of links to administrative pages were either gratuitous or as a result of collaboration.

Even though classification based on inter page relationship is more practical than classifying individual links, it is still infeasible to manually classify each page. Web pages are fewer than hyperlinks, but they cannot be efficiently classified manually because a typical UK university website contains thousands of web pages. Classification of page types can be automated using a supervised machine learning technique; decision tree induction was used in this study and results showed moderate accuracy; 71%.

Links between two staff pages suggest collaboration, and as supervised learning methods can automatically identify staff pages with decent accuracy; F measure of 77 percent, there is a possibility for more in depth analysis of inter linking between universities staffs so this type of links can be in cooperated to bibliometric analysis.

There are several other machine learning algorithms which may give more accurate results. Further work will aim to compare results of other classification techniques as well as apply natural language processing techniques to improve the accuracy of the classifier.